\documentclass[a4paper, 10pt, conference]{ieeeconf}      

\IEEEoverridecommandlockouts                              

\usepackage{multirow}

\usepackage{amsmath,amsfonts,amssymb,graphicx,bm, color}

\title{\LARGE \bf
Evaluation Uncertainty in Data-Driven Self-Driving Testing
}

\author{Zhiyuan Huang$^{1}$, 
Mansur Arief$^{2}$, Henry Lam$^{3}$, and Ding Zhao$^{2}$
\thanks{We gratefully acknowledge the support from the National Science Foundation under grants CAREER CMMI-1653339/1834710, IIS-1849280, and IIS-1849304.}
\thanks{$^{1}$Zhiyuan Huang is with the Department of Industrial and Operations Engineering at
University of Michigan, 
1205 Beal Ave, MI, USA
        {\tt\small zhyhuang@umich.edu}  }%
\thanks{$^{2}$ Mansur Arief and Ding Zhao are with the Department of Mechanical Engineering at Carnegie Mellon University, 5000 Forbes Avenue, Pittsburgh, PA, USA
        {\tt\small marief@andrew.cmu.edu
, dingzhao@cmu.edu} }%
\thanks{$^{3}$Henry Lam is with the Department of Industrial Engineering and Operations Research, Columbia University,
        500 W. 120th Street, NY, USA
        {\tt\small henry.lam@columbia.edu}}%
}

\begin{document}

\maketitle
\thispagestyle{empty}
\pagestyle{empty}

\begin{abstract}
Safety evaluation of self-driving technologies has been extensively studied. One recent approach uses Monte Carlo based evaluation to estimate the occurrence probabilities of safety-critical events as safety measures. These Monte Carlo samples are generated from stochastic input models constructed based on real-world data. In this paper, we propose an approach to assess the impact on the probability estimates from the evaluation procedures due to the estimation error caused by data variability. Our proposed method merges the classical bootstrap method for estimating input uncertainty with a likelihood ratio based scheme to reuse experiment outputs. This approach is economical and efficient in terms of implementation costs in assessing input uncertainty for the evaluation of self-driving technology. We use an example in autonomous vehicle (AV) safety evaluation to demonstrate the proposed approach as a diagnostic tool for the quality of the fitted input model. 

\end{abstract}


\section{Introduction}
The competitive race toward the mass deployment of self-driving cars driving side-by-side with human-driven vehicles on public roads advocates for an accurate and highly precise safety evaluation framework to ensure safe driving. However, achieving meaningful precision  is a challenging task when the safety-critical events under study are rare in naturalistic situations. A recent method has been developed which which adopts Monte Carlo method empowered by the Importance Sampling technique as a variance reduction scheme; this has produced appealing results. In \cite{zhao2017accelerated}, it is shown that the efficiency is enhanced by ten thousand times with the incorporation of large-scale driving data sets and the employed statistical models. This improved efficiency is highly appealing for autonomous vehicle (AV) researchers as the required testing effort is overly demanding, an estimate of 8.8 billion driving miles required to provide `sufficient' evidence to compare the safety of AV driving and human driving from logged data \cite{kalra2016driving}.

A common framework adopted to estimate the safety measure is Monte Carlo simulation, which generates a large number of samples and simulates experiments using each sample. Then an empirical expectation with confidence interval is obtained via central limit theorem. By properly integrating the Monte Carlo approach into the modeling scheme, researchers have been able to estimate the risk from driving situations based on control and trajectory prediction \cite{broadhurst2005monte, eidehall2008statistical} and corner cases in various driving scenarios \cite{zhao2017accelerated,zhao2018accelerated}. 

In the studies, the AV system is viewed as a black box and a Monte Carlo simulation is used to evaluate its safety performance. Using this view, the evaluation metric is estimated from Monte Carlo samples of variables that encompasses the uncertainties in the system, e.g. traffic environment or noise in control and observation. Usually, the stochastic models that generate the Monte Carlo samples are fitted using real-world data, and the model training is implemented separately with the safety evaluation. However, since parameters in the stochastic models are estimated from data, the variability of the data has a direct effect on the test result and the reliability of the evaluation. In this paper, this effect is referred to as \textit{input uncertainty} \cite{lam2016advanced}. In order to provide a convincing test evaluation, the input uncertainty needs to be addressed and highlighted as part of the test results. 

The goal of this paper is to provide a way to construct a confidence interval for the evaluation results as a quantitative measurement of the input uncertainty. In AV safety evaluation, since the experiment or simulation can be quite expensive, the accelerated evaluation approach studied in \cite{huang2017accelerated,zhao2017accelerated,zhao2018accelerated} was proposed to increase the efficiency of safety evaluation. Similarly, for input uncertainty quantification, an ideal method should ``minimize'' the number of experiments and computing time to be computationally economic.

In this paper, we propose an extension of the classic bootstrap technique \cite{efron1992bootstrap} for assessing the input uncertainty in Monte Carlo based evaluation approach. Our approach use a likelihood ratio based scheme for estimations in bootstrap replications, which reuses the Monte Carlo estimators for the initial evaluation of safety measures. The proposed approach significantly reduces the computation burden in the standard bootstrap implementation and enables parallel computation. Moreover, The likelihood ratio based scheme can be perfectly adapted to the accelerated evaluation approach. It can efficiently utilize the accelerated evaluation structure and provides good importance sampling estimator for each bootstrap replication in the subsequent stages.

The remainder of the paper is structured as follows: Section \ref{sec:prob_Setting} introduces the notations and sets up the problem. Section \ref{sec:method} reviews classic bootstrap schemes and presents the proposed approach. Section \ref{sec:numerics} illustrates the implementation details in the proposed approach using numerical examples and demonstrates using an AV evaluation example. The conclusion is provided in Section \ref{ref:conclusion}.

\section{Problem Setting Under AV Testing}\label{sec:prob_Setting}
In this section, we set up the notations for the AV evaluation problem. Then we will analyze the influence of input uncertainty in the evaluation and the potential issues that may arise if the input uncertainty is ignored.
 
\subsection{AV Safety Evaluation Setting}
The goal of the Monte Carlo based test approach is to understand the performance of the AV system under uncertain environment. We use $\xi\in \mathbb{R}^d$ to denote $d$-dimensional uncertain factors to the AV system, where each element represents one attribute of the environment. We use $\xi_i$ to denote a realization of the random vector $\xi$. In Accelerated Evaluation approaches \cite{zhao2017accelerated,zhao2018accelerated, huang2017accelerated}, the traffic environment is considered an uncertainty for the AV system under study and is represented by $\xi$. 

A parametric stochastic model $p(\xi|\theta)$ is used to represent the uncertainty, where $\theta \in \mathbb{R}^m$ represents the parameters in parametric stochastic model. The model can be static  \cite{huang2017accelerated,zhao2017accelerated}, or dynamic, which represents stochastic processes \cite{zhao2018accelerated}. The underlying assumption is that the parametric model contains the true distribution. In another word, there exists a parameter $\theta_0$ such that $\xi \sim p(\xi|\theta_0)$.

Here, we use $f(\xi)$ to denote the performance measurement of an AV system under environment $\xi$, which is referred to as \textit{performance function}. For instance, we are interested in a certain type of safety-critical event (e.g. a crash) that occurs to the AV system under the environment $\xi$, and we use $\varepsilon \subseteq \mathbb{R}^d$ to denote the set of safety-critical event. Then the performance function is defined as $f(\xi)=I_\varepsilon(\xi) \in \{0,1\}$, which indicates whether a certain type of safety-critical event (e.g. a crash) is occurred to the AV system under the environment $\xi$, where 0 and 1 represent whether a safety-critical event occurring (1 for crash). For example, $f(\xi)$ can be the test results of a trail in computer simulation, a real-world on-track trail, or a trail in a particular zone of a public street. Therefore it can be rather expensive to run experiment trail for $f(\xi)$.

Our goal is to estimate the expectation of $f$ given by
\begin{equation}\label{eq:expecation}
    E[f(\xi)|\theta_0]=\int_\xi f(\xi) p(\xi|\theta_0) d \xi .
\end{equation} In Accelerated Evaluation, this expectation is the probability of the safety critical event, which is revealed by the equality 
\begin{equation}
    E[f(\xi)|\theta_0]=E[I_\varepsilon(\xi)|\theta_0]=P(\xi \in \varepsilon).
\end{equation} 
This measure is used as the criterion for the safety of a tested AV, which is denoted by $\gamma$ in later discussion.

Usually the performance function is defined by complex systems, and the expectation \eqref{eq:expecation} is hard to be analytically computed even if $p(\xi|\theta_0)$ is fully known. Hence the Monte Carlo approach is applied to estimate $E[f(\xi)|\theta_0]$. For crude Monte Carlo, we generate samples $\xi_1,...,\xi_n$ from $p(\xi|\theta_0)$ and estimate $E[f(\xi)|\theta_0]$ using the sample mean 
\begin{equation}
    \hat {E}[f(\xi)|\theta_0] = \frac {\sum_{i=1}^n f(\xi_i)}{n}.
\end{equation}
Each evaluation of the performance function $f(\xi_i)$ at a certain sample $\xi_i$ is referred to as one \textit{experiment trail}. 

In the context of AV testing, we expect the safety-critical event to be very rare ($\gamma < 10^{-5}$), where crude Monte Carlo is inefficient. The inefficiency is reflected in the large relative error (${error}/p$) of the crude Monte Carlo estimator. To intuitively explained this, we can consider that every $\xi$ drawn from $p(\xi,\theta_0)$ returns $f(\xi)=0$ with probability $1-p$, and therefore huge number of samples are required to obtain a safety-critical event. The computation cost is usually prohibitive for obtaining an accurate estimation (in terms of relative error) due to expensive experiment trials.

To improve the efficiency in estimating $E[f(\xi)|\theta_0]$, \cite{zhao2017accelerated} uses importance sampling estimator to reduce the variance. Instead of drawing samples from $p(\xi|\theta_0)$, we construct an accelerating distribution $\tilde{p}(\xi)$ based on information of $p(\xi|\theta_0)$ and $\varepsilon$. With samples $\xi_1,...,\xi_n$ from $\tilde{p}(\xi)$, we use the estimator 
\begin{equation}
\label{eq:IS_estimator}
    g(\xi_i,\theta_0)=\frac{p(\xi_i|\theta_0)}{\tilde{p}(\xi_i)}f(\xi_i),
\end{equation} 
which can be proved to be unbiased. With a good selection of $\tilde{p}$, the importance sampling estimator can be very efficient. \cite{huang2017accelerated} has shown that the importance sampling estimator can achieve the same accuracy as the crude Monte Carlo estimator using only $10^{-3}$ of the crude Monte Carlo samples. Then we use the sample mean \begin{equation}
    \bar{g}_0=\frac{\sum_{i=1}^{n} g(\xi_i,\theta_0)}{n}
\end{equation}
to estimate the expectation $E[f(\xi)|\theta_0]$.

Usually a confidence interval is constructed as a reference of the accuracy of the estimation. For a confidence interval with confidence level $1-\alpha$, we want to have
\begin{equation}
    P(E[f(\xi)|\theta_0]\in \left[ C_L, C_U \right] )\geq 1-\alpha,
\end{equation}
i.e. we want the confidence interval $ \left[ C_L, C_U \right]$ to cover the truth with probability greater than $1-\alpha$. The most commonly used confidence interval for sample mean is derived from central limit theorem \cite{asmussen2007stochastic}, which uses
\begin{equation}
\label{eq:ci_sim_l}
    C_L=\bar{g}_0 - z_{\alpha/2} \sqrt{ \hat{Var}_{\xi} \left( \bar{g}_0\right)}
\end{equation} and 
\begin{equation}
\label{eq:ci_sim_u}
    C_U=\bar{g}_0 + z_{1-\alpha/2} \sqrt{ \hat {Var}_{\xi} \left( \bar{g}_0\right)},
\end{equation}
where $z_\alpha$ denotes the $\alpha$ quantile of standard Gaussian distribution.

\subsection{Input Uncertainty in Safety Evaluation}

In this paper, we consider the situation where $\theta_0$ is unknown but a finite number of data from $p(\xi|\theta_0)$ is available. We use the maximum likelihood estimation (MLE) $\hat{\theta}$ for the parameter $\theta_0$ in the stochastic model. Note that $\hat{\theta}$ is a consistent estimator of $\theta_0$, i.e. 
$\hat \theta$ converges to  $\theta_0$ in probability as the number of samples increases.
Since $\hat{\theta}$ is estimated from data of $\xi$, it is random due to the variability of the samples. 

In practice, if the environment modeling and the evaluation are implemented as separate tasks, the estimated parameter $\hat \theta$ will be used as the true parameter. That is, the estimator is given by \begin{equation} \label{eq:estimator_for_performance}
    \Bar{g}=\frac {\sum_{i=1}^n g(\xi_i,\hat{\theta})}{n},
\end{equation} where $\xi_i$'s are generated from $p(\xi|\hat \theta)$. The confidence interval from \eqref{eq:ci_sim_l} and \eqref{eq:ci_sim_u} uses the variance estimated from the samples, that is  \begin{equation}
    \frac{\sum_{i=1}^n (g(\xi_i,\hat{\theta})- \Bar{g})^2}{n-1}.
\end{equation} 

However, the above approach ignores the variation sourced from the estimated parameter $\hat \theta$, where we consider as the input uncertainty. The influence of input uncertainty can be revealed by a decomposition of the variance of $\Bar{g}$: \begin{equation}\label{eq:decomposition_uncertainty}
    Var(\Bar{g})=Var_{\hat{\theta}} \left( E_{\xi} \left[ \Bar{g}| \hat{\theta}\right] \right) +E_{\hat{\theta}} \left[ Var_{\xi} \left( \Bar{g}| \hat{\theta}\right) \right].
\end{equation}
In this decomposition, the first term is the input uncertainty and the second term is referred to as \textit{simulation uncertainty}. We note that if we ignore the variation of $\hat{\theta}$, only $ Var_{\xi} \left( \Bar{Y}| \hat{\theta}\right)$ would be considered as the variance of the estimator. The resulted confidence interval will be likely to undercover for the truth $E[f(\xi)|\theta_0]$. Under the AV evaluation context, confidence intervals that undercover for the truth are harmful for the reliability of the evaluation. 

In this paper, our goal is to construct confidence intervals that target to cover $E[f(\xi)|\theta_0]$ with confidence level $1-\alpha$. This is a way to quantify the input uncertainty and therefore provide an assessment of the reliability of the evaluation results.

\section{Measurement of Input Uncertainty}\label{sec:method}

In this section, we first introduce some well-studied bootstrap framework. We then propose our approach based on these techniques. 

\subsection{Classic Bootstrap Approach}\label{sec:classic_boot}
The bootstrap technique dates back to \cite{efron1992bootstrap,efron1982jackknife}, which is studied to estimate the variability of statistical estimators by judiciously reusing the data. \cite{cheng1997sensitivity} considers a parametric version of bootstrap for assessing the input uncertainty in simulation. For further interests of input uncertainty quantification, one can refer to \cite{barton2002panel,henderson2003input,chick2006bayesian,barton2012input,song2014advanced,lam2016advanced} and \cite{nelson2013foundations}, Section 7.2.

We first clarify some notations to avoid confusion. Note that the random vector $\xi$ and its samples appears in both the input modeling part and the simulation part. We use $X_i$'s to denote samples that we collected from the real world and used to estimate $\theta$. We use $\xi_i$'s to represent the samples in the simulation part, which are generated from a certain distribution $\tilde{p}$ and are used to evaluate the estimator $Y(\xi_i,\theta)$.  

In general, a bootstrap scheme for quantifying input uncertainty is as follows. We first generate samples $\hat \theta^1,...,\hat \theta^B$ that approximately follows the true distribution of $\hat \theta$. For each $\hat \theta_i$, we generate samples $\xi_1,...,\xi_r$ from $p(\xi;\theta)$ and estimate $\bar{g}^i$ using 
\begin{equation}
\label{eq:classic_boot_est}
    \bar{g}^i=\frac{1}{r} \sum_{j=1}^r g(\xi_j,\hat \theta^i).
\end{equation}
After computing $\bar{g}^1,...,\bar{g}^B$, we find the $\alpha/2$ and $1-\alpha/2$ - the empirical quantiles of $\bar{g}^1,...,\bar{g}^B$ as lower and upper bound of the confidence interval, respectively. We denote as $C_L=\hat q_{\alpha/2}(\bar{g}^i)$ the lower bound and as $C_U=\hat q_{1-\alpha/2}(\bar{g}^i)$ the upper bound. We list three bootstrap approaches in the Appendix.

\subsection{The Proposed Approach: A Likelihood Ratio Based Estimation for Bootstrap}

To motivate the proposed approach, we first consider the computation cost for a classic bootstrap scheme. No matter what bootstrap scheme we use, after we obtain the bootstrapped parameters $\hat{\theta}^1,...,\hat{\theta}^B$, we would need to estimate $E[g|\hat{\theta}^i]$ using $\bar{g}^i$. To obtain a good empirical quantile, we usually require $B$ to be as large as possible (usually hundreds or more)\cite{efron1992bootstrap}. Also, in order to reduce the simulation uncertainty to avoid obtaining an over-covered confidence interval, we want $r$ to be as large as possible. The number of experiment trials in total will be $rB$, which is $B$ times more than estimating the probability. When the experiment is expensive and time-consuming, the price for assessing the input uncertainty might not be affordable. Here, we propose an approach that can assess the input uncertainty with no additional cost for experiment trials.

Assume we have already estimated the average performance measure from samples $\xi_1,...,\xi_n$ from $\tilde{p}(\xi)$ using 
\eqref{eq:estimator_for_performance}, where $\tilde{p}(\xi)$ can be $p(\xi,\hat \theta)$ or an appropriate accelerating distribution for $p(\xi,\hat\theta)$. Then, we obtain bootstrap parameters $\hat \theta^1,...,\hat \theta^B$ using any bootstrap scheme. For each $\hat \theta^i$, instead of generate a new sample from $p(\xi,\hat \theta^i)$, we use the same set of samples $\xi_1,...,\xi_n$, and estimate $\bar{g}^i$ using \begin{equation}
    \bar{g}^i= \frac{1}{n}\sum_{j=1}^n \frac{p(\xi_j,\hat\theta^i)}{p(\xi_j,\hat\theta)}g(\xi_j,\hat{\theta}),
\end{equation} 
We should note that each $\bar{g}^i$ is still an unbiased estimator, i.e. we have
\begin{equation}
    E \left[\frac{p(\xi,\hat\theta^i)}{p(\xi,\hat\theta)}g(\xi,\hat{\theta})\right]=E\left[g(\xi,\hat\theta^i)\right].
\end{equation}
Note that by estimating $\bar g _i$ in this way, we do not need to evaluate $f(\xi)$ (which is hidden in $g$) at any new realization of $\xi$.

This approach can fit into the accelerated evaluation framework, i.e. when $\tilde p$ is a good accelerated distribution for $p(\xi,\hat\theta)$ and $g$ is defined by \eqref{eq:IS_estimator}. By using the likelihood ratio adjustment, the samples across different resampled value of $\hat \theta^1,...,\hat \theta^B$ are now correlated. It is unclear how this would affect the reliability of our estimate, but this issue would likely go away when $r$ is large enough (which is the case in accelerated evaluation). Secondly, the likelihood ratio adjustment can sometimes blow up the magnitude of the output estimate, especially when the when the estimation of $\hat \theta$ is highly uncertain. Since this is also a sign that the stochastic model is unreliable, this issue would not affect the use of the proposed approach. Usually we can speculate the $\tilde{p}(\xi)$ would also be a good accelerating distribution for $p(\xi,\hat{\theta}^i)$. For instance, if we use exponential tilting of Exponential distribution, the optimal $\tilde{p}(\xi)$ for a certain performance function $f(x)$ is the same for any parameter values $\theta$ for the exponential distribution. By using the proposed approach, we saved $r(B-1)$ experiment trials compared to the classical bootstrap approaches.

\section{Numerical Experiments and Discussion}\label{sec:numerics}
In this section, we present some numerical experiments to illustrate the proposed approach and discuss some implementation details. We first discuss the performance of the three bootstrap schemes under different scenarios. We then use a simple illustrative problem to demonstrate the proposed approach. Lastly, we apply the proposed approach on an AV testing example problem.

\subsection{Comparison of Bootstrap Schemes}
In Appendix, we introduce three bootstrap schemes that are applicable to our framework. Here we use some numerical studies to show the advantages of each scheme and provide a guideline of choosing suitable scheme in different conditions.

The purpose of the experiment is to investigate if $\hat \theta^i$'s generated using these bootstrap schemes are roughly close to the true distribution of $\hat \theta$ with different numbers of samples $k$. In the experiment, we first generate $k$ samples from $p(\xi|\theta_0)$. For each bootstrap scheme, we use these samples to generate $\hat \theta^1,...,\hat \theta_B$ with $B=1000$. We use the $\alpha/2$ and $1-\alpha/2$ empirical quantile of these $\hat \theta^i$'s as upper and lower bound for a confidence interval and check whether $\theta_0$ is covered. We repeat this procedure for 1000 times with an independently generated sample set. We use the coverage of the truth to test the accuracy of the confidence interval obtained from these schemes. We use $\alpha=0.05$ in our experiments.

The experiment results with different $k$ and different distribution models are tabulated in Tables \ref{table:para_exp}. In the table, ``Plain'' represents plain bootstrap, ``Parametric'' represents parametric bootstrap, ``Asym Cls'' stands for the asymptotic distribution scheme using closed form Fisher's information and ``Asym Est'' stands for the asymptotic distribution scheme using empirical Fisher's information. See Appendix for details of these methods.

\begin{table}[t]
\centering
\caption{The CI coverage of true parameter $\mu$ in exponential distribution using three bootstrap schemes.}
\label{table:para_exp}
\begin{tabular}{|l|l|l|l|l|}
\hline
Samples        & Approach                   & Object & Coverage \\ \hline
\multirow{4}{*}{k=10} & Plain                     & $\mu$        & 84.70\%    \\ \cline{2-4} 
                   &     Parametric           & $\mu$      & 92.20\%     \\ \cline{2-4}  
                   &    Asym Cls              & $\mu$        & 88.30\%        \\ \cline{2-4}  
                   &    Asym Est          & $\mu$    & 90.20\%       \\ \hline
\multirow{4}{*}{k=20} & Plain                     & $\mu$        & 91.40\%   \\ \cline{2-4}  
                   &     Parametric           & $\mu$      & 93.10\%    \\ \cline{2-4}  
                   &    Asym Cls              & $\mu$        & 93.30\%        \\ \cline{2-4}  
                   &    Asym Est          & $\mu$    & 92.00\%         \\ \hline      
\multirow{4}{*}{k=100} & Plain                     & $\mu$        & 94.10\%  \\ \cline{2-4}  
                   &     Parametric           & $\mu$      & 95.10\%    \\ \cline{2-4} 
                   &    Asym Cls              & $\mu$        & 94.30\%        \\ \cline{2-4} 
                   &    Asym Est          & $\mu$    & 95.20\%         \\ \hline 
\end{tabular}
\end{table}

We consider exponential distribution for $p(X|\theta_0)$. From Table \ref{table:para_exp}, we observe that when $k=10$, the coverage rates for all schemes have an obvious gap to the target 95\%. For the plain bootstrap, the relatively low performance is due to the very small sample size used to construct the empirical distribution. For the parametric bootstrap, this is caused by the error in estimating $\hat \theta$. The asymptotic approaches suffer from both bad estimation of $\hat \theta$ and small value of $k$ (note that asymptotic analyses require $k \rightarrow \infty$). Among these approaches, the parametric bootstrap has the smallest gap. This is partly because the assumption of the correct parametric model remedies the error from the variability of the samples. As we increase the value of $k$, the gap between target coverage and the obtained coverage reduces. When we use $k=100$, the coverage rates for all schemes are already close to the target.

In summary, the parametric bootstrap provides a better coverage of the truth. especially when the number of samples is very small. The coverage for these schemes is similar when the number of samples is large enough. In a sufficient sample size situation, the asymptotic schemes have an upper hand for the efficiency of generating the parameters.

\subsection{Illustrative Problem}
We consider a simple probability estimation problem to demonstrate the effectiveness of the proposed approach in providing a valid confidence interval. This is shown in two aspects: a) the coverage of the proposed approach is close to the target; b) the confidence interval width is relatively narrow.

\begin{table}[]
\centering
\caption{The coverage and average width of confidence intervals from various approaches}
\label{table:prob_boot_gaus}
\begin{tabular}{|l|l|l|l|l|l|l|}
\hline
Samples & 100 &1000  &10000   \\ \hline
Coverage CF & 0.9432 &0.9451  &0.9505 \\ \hline
CI Width CF & 1.33e-05 &8.85e-07  &2.20e-07\\ \hline
Coverage LR & 0.9426 &0.9444  &0.9486\\ \hline
CI Width LR & 1.33e-05 &8.85e-07  &2.20e-07 \\ \hline
Coverage SU & 0.0177 &0.0630  &0.1903\\ \hline
CI Width SU & 8.28e-08 &3.08e-08  &2.72e-08\\ \hline
\end{tabular}
\\
\vspace{1ex}
{\raggedright \hspace{9ex} CF (closed form), LR (likelihood ratio), \\
\hspace{9ex} SU (simulation uncertainty only)\par}
\end{table}

We consider estimating the probability of $P(\xi > \beta)$, where $\xi$ follows a standard Gaussian distribution and we use $\beta=5$. The choice of the problem is because we have an analytic solution for the probability, which makes it easier to validate whether the constructed confidence interval covers the truth or not. We use different numbers of sample size $k$ for estimating $\hat \theta =\{\hat \mu,\hat \sigma\}$. We use $B=1000$ bootstrap samples for constructing the confidence interval. For the estimation of $\bar g^i$, we consider two approaches. The first is to use the proposed approach with 10,000 importance sampling estimators. To show the constructed CI has a relative narrow width, we also consider using the analytic solution for $P(\xi > \beta)$ for each $\hat \theta^i$ as a baseline (so that there is no simulation uncertainty). We repeat for 10000 total replications and compute the coverage of the confidence interval. 

The experiment results are summarized in Table \ref{table:prob_boot_gaus}, where we show the coverage and width of the confidence intervals computed using the closed form probability (CF), likelihood ratio (LR) estimation, and Equation \eqref{eq:ci_sim_l} and \eqref{eq:ci_sim_u} that only consider simulation uncertainty (SU), i.e. with input uncertainty ignored. 

We have two main observations from these experiment results. First, the proposed likelihood ratio scheme provides a good estimation of the probability of interest. This claim is supported by the similar coverage rates and confidence interval width for the closed form baseline approach and our proposed approach.  Second, the confidence interval computed without incorporating the input uncertainty is problematic. This observation is revealed by the low coverage rates (especially when the estimator has high variability, e.g. when the sample size is small) and narrow confidence interval width. 

\subsection{Accelerated Evaluation Example}

\begin{figure}[!htb]
     \centering
     {\includegraphics[width=0.4\textwidth]{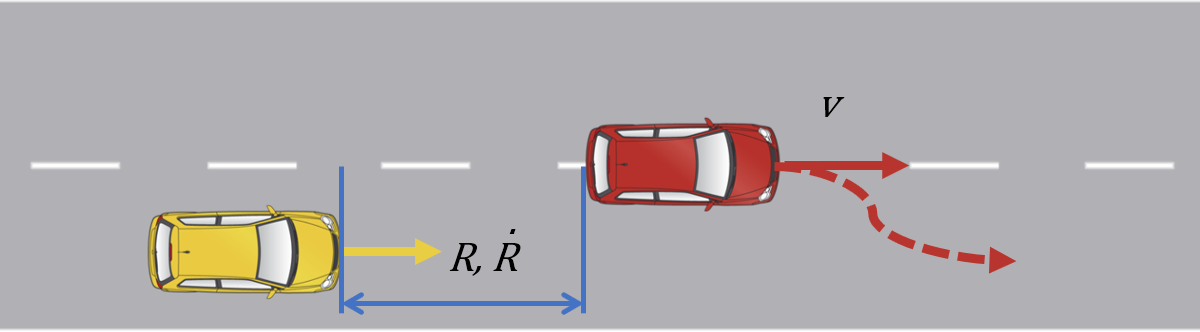}} 
     \caption{An illustration of the lane change scenario in AV evaluation.}
     \label{fig:lane_change}
 \end{figure}

To demonstrate the proposed approach, we consider the AV evaluation problem and AV model discussed in \cite{zhao2017accelerated}. The lane change test scenario is shown in Figure \ref{fig:lane_change}, where we evaluate the safety level of a test AV by estimating the probability of crash when a frontal car cut into the lane. Crash is determined by whether the minimum range of two vehicles during the lane change procedure reaches 0. The traffic environment in this scenario is represent by $v$, the initial velocity of the frontal vehicle, $R$, the initial range between the two vehicles, and $TTC$, the time-to-collision value defined by $TTC=R/\dot{R}$.

In our problem, we consider the frontal car to have an initial velocity $v=30m/s$, which is a common speed in highway driving. We extract 12,304 lane change scenario samples identified from the SPMD dataset \cite{Bezzina2014} with similar velocity. We use the samples to fit $R^{-1}$ and $TTC^{-1}$ with exponential distribution. We used the cross-entropy method to find an optimal accelerating distribution by exponential tilting for $R^{-1}$ and $TTC^{-1}$; we generated $10^5$ samples from the accelerating distribution. We then use the proposed approach to construct confidence interval that incorporates the input uncertainty. 

  \begin{figure}[!htb]
     \centering
     {\includegraphics[width=0.4\textwidth]{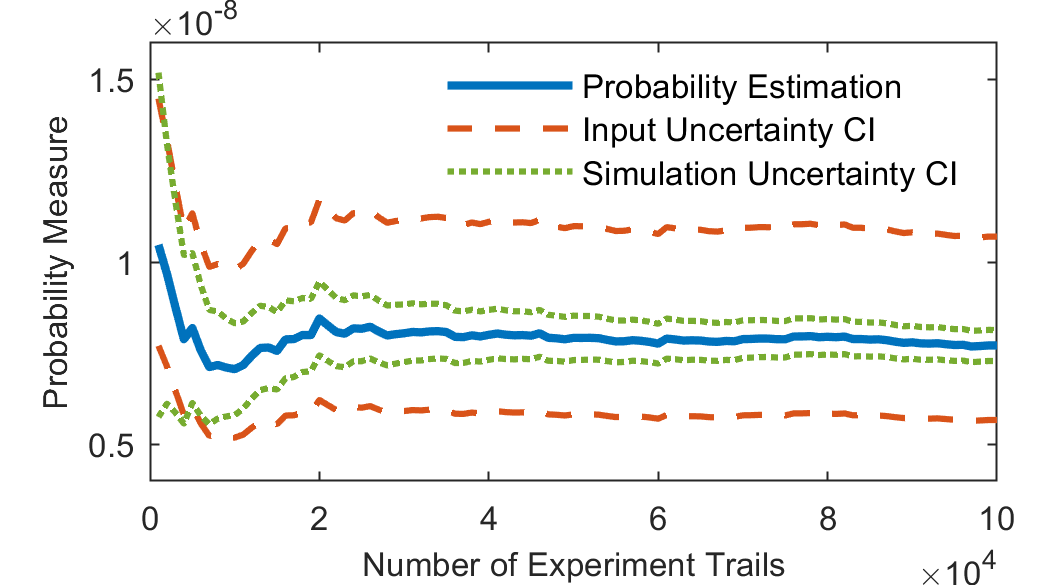}} 
     \caption{The estimates of safety critical events rate (probability of crash) and their confidence intervals with different number of experiments.}
     \label{fig:prob_Est}
 \end{figure}
 
 In Figure \ref{fig:prob_Est}, we present the probability estimation and the two types of confidence intervals we construct given different number of samples. The confidence interval for simulation uncertainty is estimated using \eqref{eq:ci_sim_l} and \eqref{eq:ci_sim_u}. We observe that the confidence interval for simulation uncertainty has a much smaller width than the input uncertainty width. This observation indicates that if the input uncertainty is ignored, the evaluation results can be misleading. For instance, if we use the confidence upper bound to interpret the safety level of a vehicle, the input uncertainty upper bound is roughly 1.5 times of the simulation uncertainty, hence using only the simulation uncertainty would underestimate the risk of crash. 
 
  \begin{figure}[!htb]
     \centering
     {\includegraphics[width=0.4\textwidth]{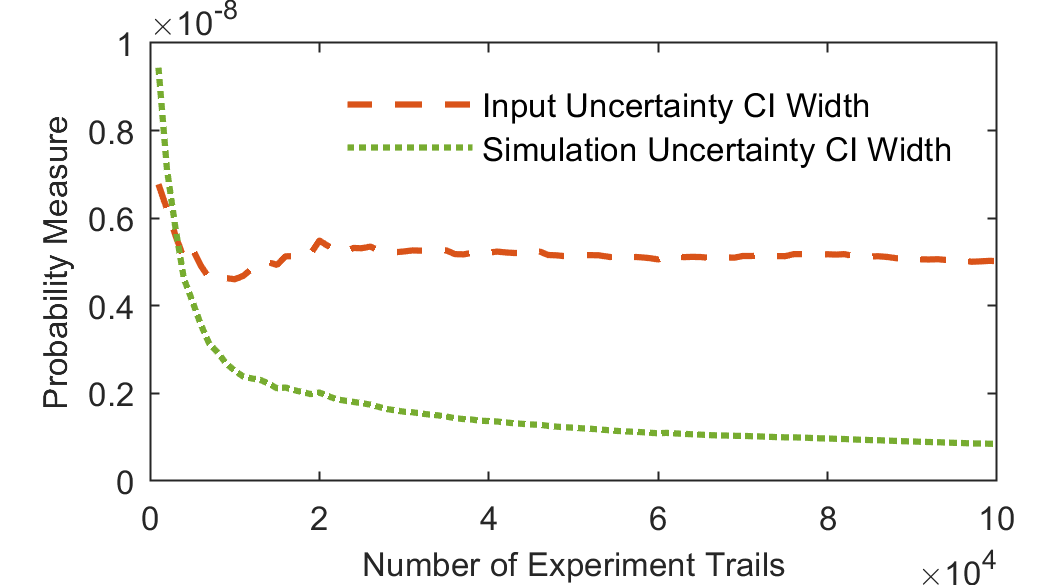}} 
     \caption{The width of confidence intervals for estimates of safety critical events rate (probability of crash) with different number of experiments.}
     \label{fig:ci_width}
 \end{figure}
 
 Figure \ref{fig:ci_width} shows how the widths of the two intervals changes as the number of experiment trials increases. As known in literature, the width of the simulation uncertainty confidence interval shrinks in the order of $O(1/\sqrt{n})$. This trend can be easily observed from the figure. On the other hand, since we are not changing the number of samples we use to estimate $\hat \theta$, the input uncertainty confidence interval should not change as $n$ increases, which is confirmed by the figure. When the number of experiment trials is sufficiently large, the simulation uncertainty decreases while the input uncertainty remains the same as the number of trials increases. When $n$ is small, the CI width of input uncertainty is not as stable as when $n$ is large. This is because when we do not have enough samples, the simulation uncertainty becomes large and can perturb the estimation of the input uncertainty. 
 
 This implies that input uncertainty can partially help reveal the consistency of information from the data and  ignoring it could lead to misleading results and wrong conclusion. In the case where one can ensure the $n$ data provide good representativeness for the whole unseen data, input uncertainty analysis help describe the information richness of the collected data with regard to the model and evaluation results, which are essential for rigorous AV evaluation purposes.

 \section{Conclusion}
 \label{ref:conclusion}
 In this paper, we propose an approach to assess the input uncertainty in Monte Carlo- based AV testing methods, which requires zero additional experiment trails. The proposed approach is shown to be computationally efficient and easy to implement while providing valid confidence intervals that incorporate input uncertainty. In the future, we would consider extending our study to model-free input uncertainty analysis for a wider application domains.

\section*{APPENDIX}

\subsection{Examples of Bootstrap Methods} \label{sec:bootstrap}

Here, we introduce three different schemes for generate samples $\hat \theta^1,...,\hat \theta^B$ that are straight-forward and easy to implement. For a sound empirical study on the performance of bootstrap schemes, refer to \cite{barton2010framework}. 

These bootstrap schemes assume that we start with a sample set $\{X_1,...,X_k\}$ and the MLE $\hat \theta$ is estimated using these samples. Note that in the discussed approaches, we restrict the resampling size to be equal to the original sample size, namely $k$. This is not required for the bootstrap technique, but we adopt this setting for convenience and simplicity.

\subsubsection{Plain Bootstrap} 
Plain bootstrap considers the sample set $\{X_1,...,X_k\}$ as an empirical distribution, say $\hat{f}$, and use it as an approximation of the real distribution of $X_i$. We draw $k$ samples from $\hat{f}$, i.e. resample from $\{X_1,...,X_k\}$ with replacement, and then use these samples to estimate $\hat{\theta}^1$ (using MLE). We repeat this procedure for $B$ times to obtain $\hat{\theta}^1,...,\hat{\theta}^B$.

\subsubsection{Parametric Bootstrap}
Here we use $p(X,\hat{\theta})$ as an approximation of the real distribution of $X_i$. We draw $k$ samples from $p(X,\hat{\theta})$ and use them to estimate $\hat{\theta}^1$. We repeat this for $B$ times and collect $\hat{\theta}^i,\ i=1,...,B$.

\subsubsection{Sample Parameters from Asymptotic Distribution}
Since the $\hat{\theta}$ is estimated using MLE, we know the asymptotic behavior of $
\hat \theta$. That is when $k\rightarrow \infty$, we have 
\begin{equation}
    \sqrt{k}(\hat{\theta}-\theta_0)\sim N(0, I^{-1}(\theta_0)),
\end{equation} 
where $I^{-1}(\theta)$ is the inverse of Fisher's information matrix of the parametric distribution $p(\cdot|\theta)$. Since $\theta_0$ is unknown, we can use its MLE $\hat \theta$ to obtain an approximation of the asymptotic distribution $N(0, I^{-1}(\hat \theta))$. 

In practice, one can use the empirical Fisher's information matrix, which is an estimation based on the samples. That is \begin{equation}
    \hat I(\theta)= - \frac{1}{k} \sum_{i=1}^k  \frac{\partial^2}{\partial \theta^2}  \log p(X_i|\theta),
\end{equation}
where $X_i$'s are the samples we use to fit the model. Thus, we can direct sample $\hat{\theta}^1,...,\hat{\theta}^B$ from $N(\hat{\theta},I^{-1}(\hat{\theta})/k)$ or $N(\hat{\theta},\hat{I}^{-1}(\hat{\theta})/k)$, which reduces computation cost from resampling and estimating $\hat{\theta}^i$.



\bibliographystyle{IEEEtran}
\bibliography{citation.bib}

\end{document}